\documentclass[letter,tradisabstract]{aa}  

\usepackage{graphicx}
\usepackage{bm}
\usepackage{txfonts}
\newcommand\Mout{\dot{M}_\mathrm{inf}}
\newcommand\Lout{\mathcal{L}_\mathrm{out}}
\newcommand\tinj{\theta_\mathrm{inj}}
\newcommand\Rout{R_\mathrm{out}}
\newcommand\td{T_{100}}
\newcommand\Rd{R_\mathrm{d}}

\usepackage{color}
\usepackage{amstext}

\begin{document}

   \title{Spiral-driven accretion in protoplanetary discs}

   \subtitle{I. 2D models}

   \author{Geoffroy Lesur
          \inst{1,2}
          \and
          Patrick Hennebelle\inst{3}
          \and
          S\'ebastien Fromang\inst{3}
          }

   \institute{Univ. Grenoble Alpes, IPAG, F-38000 Grenoble, France\\
   \email{geoffroy.lesur@ujf-grenoble.fr}
   \and 
                CNRS, IPAG, F-38000 Grenoble, France
             \and
             Laboratoire AIM, CEA/DSM--CNRS--Universit\'e Paris 7, Irfu/Service d'Astrophysique, CEA-Saclay, 91191 Gif-sur-Yvette, France
             }

   \date{Received 11/06/2015}

 
  \abstract{We numerically investigate the dynamics of a 2D non-magnetised protoplanetary disc surrounded by an inflow coming from an external envelope. We find that the accretion shock between the disc and the inflow is unstable, leading to the generation of large-amplitude spiral density waves. These spiral waves propagate over long distances, down to radii at least ten times smaller than the accretion shock radius. We measure spiral-driven outward angular momentum transport with $10^{-4}\lesssim \alpha< 10^{-2}$ for
an inflow accretion rate $\Mout\gtrsim 10^{-8}\,M_\odot\cdot\mathrm{yr}^{-1}$. We conclude that the interaction of the disc with its envelope leads to long-lived spiral density waves and radial angular momentum transport with rates that cannot be neglected in young non-magnetised protostellar discs.}


   \maketitle
%

\section{Introduction}
Accretion is an essential phenomenon in astrophysics as it is the way through which gravitational energy is transformed into heat and therefore observable radiation. However, the physical process responsible for accretion is still debated. Although the magnetorotational instability (MRI, \citealt{BH91}) provides an efficient accretion mechanism, its applicability to protoplanetary discs is questionable since these objects are very weakly ionised and might quench the MRI through various non-ideal magnetohydrodynamical (MHD) processes \citep{TF14}. 

The usual approach to protoplanetary disc dynamical modelling is to assume that these objects are isolated in space. In this context, most of the models rely on processes such as MRI-driven turbulence, self-gravity, or winds to drive accretion. However, these discs are not isolated systems, and the question is whether surrounding material, albeit much less dense than the disc, could perturb it sufficiently to drive accretion. This possibility was proposed by \cite{PK05} to explain accretion rates that scale like the central protostar mass squared. This scenario was later refined by \cite{TB08}, \cite{KH10}, and \cite{PH14} and interpreted as a consequence of Bondi-Hoyle accretion onto the protostar.

In this letter, we explore the importance of external accretion using a very simplified model. We consider a 2D hydrodynamic Keplerian disc onto which falls gas coming from a surrounding envelope. The disc is inviscid and hydrodynamically stable so that no accretion can occur without this external inflow. We first present in detail the model and the numerical method used to solve the equations of motion. We then explore some of the results obtained using this model and finally discuss the limitations and implications of our findings.

This work shares some similarities with that of \cite{BHZ15}: both consider accretion discs subject to an external inflow using 2D hydrodynamical models. However, we focus here on the transport of angular momentum far from the perturbed region using extended disc models, whereas \cite{BHZ15} concentrated on the local response of the disc in the presence of viscous angular momentum transport. We discuss the differences between the outcome of these models in Sect. 4.


\section{Model}
\subsection{Physical model}
We study the dynamics of an accretion disc close to Keplerian rotation. We consider a 2D frame $(R,\theta)$ and integrate the equations of motion in the vertical direction. In this initial study, we neglect the role played by magnetic fields and self-gravity to concentrate on a purely hydrodynamical model. The equations of motion read
\begin{align}
\label{eq:cons}\frac{\partial \Sigma}{\partial t}+\bm{\nabla\cdot}(\Sigma \bm{v})&=0,\\
\label{eq:mom}
\frac{\partial \bm{v}}{\partial t}+\bm{v\cdot\nabla v}&=-\frac{1}{\Sigma}\bm{\nabla}P-\frac{GM_\odot}{R^2},
\end{align}
where $\bm{v}$ is the flow velocity, $\Sigma$ the surface density, $P$ the vertically averaged pressure, $G$ the gravitational constant, and $M_\odot$ the mass of the central object, which is assumed to be a solar-mass star. 

Instead of the usual minimum mass solar nebula (MMSN) density profile, we use a power-law surface density profile with an exponential outer cut-off
\begin{align}
\label{eq:density_profile}
\Sigma=\Sigma_0 \Bigg(\frac{R}{1\,\mathrm{a.u.}}\Bigg)^{-1}\exp\Bigg(-\frac{R}{R_c}\Bigg)\,,
\end{align}
where we choose $\Sigma_0=1700\,\mathrm{g.cm}^{-2}$ and $R_c=40\,\mathrm{a.u.,}$ which corresponds to typical protoplanetary discs observed in Ophiuchus \citep{AW09} with a mass of $4.8\times 10^{-2}\,M_\odot$. The temperature profile is chosen to be locally isothermal so that $c_s/v_\mathrm{K}\equiv\varepsilon=0.1$ or $0.05,$ where $c_s$ is the sound speed and $v_\mathrm{K}\equiv\sqrt{G M_\odot/R}$ is the Keplerian velocity. The equation of state is then simply $P=c_s^2\Sigma$. Assuming a vertically isothermal hydrostatic profile, this model leads to an accretion disc with a constant aspect ratio $H/R=\varepsilon,$ where $H$ is the typical vertical disc scale height.

\subsection{Boundary conditions}
The outer radial boundary condition is located at $\Rout=400\,\mathrm{a.u}$.
To mimic matter falling onto the disc, we inject material at the outer radial boundary condition that then falls onto the disc. The material is injected over an azimuthal extent $0<\theta < \tinj$ with a sonic radial velocity.  $v_{r_\mathrm{out}}=-c_s$. 
We vary three parameters for the injected material: its specific angular momentum $\Lout\equiv v_\theta(R_\mathrm{out})R_\mathrm{out}$, its accretion rate $\Mout\equiv- \int \mathrm{d}\theta \,\Sigma_\mathrm{out}v_{r_\mathrm{out}}$, $\Sigma_\mathrm{out}$ being deduced from the desired accretion rate, and the azimuthal width of the inflow $\tinj$.

The inner radial boundary is located at $R_\mathrm{in}=1\,\mathrm{a.u}$. The inner boundary condition is forced to the initial density and velocity profile. To avoid spurious reflexion of spiral density waves, we add a damping zone in the region $1\,\mathrm{a.u}<R<3\,\mathrm{a.u}$. In this region, we exponentially relax non-axisymmetric velocity fluctuations on a fixed timescale set to eight local orbital periods. To avoid mass accumulation in the damping zone, we also relax the density to the initial density profile on the same timescale. 

\subsection{Numerical algorithm}
We use Pluto 4.0 \citep{M07} to solve the equations of motion on a cylindrical grid, using log-spaced grid cells in the radial direction. Parabolic reconstruction is used in each cell to cope with the non-uniform radial grid, and a third-order  Runge-Kutta algorithm is used to evolve the system in time. We use the HLL Riemann solver\footnote{We also ran simulations with the HLLC Riemann solver. In this case, however, the noise level of the background spirals in the absence of inflow is $\alpha\sim 10^{-4}$ because there is almost no numerical dissipation. We therefore only present results obtained with HLL for which the background noise level is much lower.}  at cell boundaries combined with orbital advection \citep{M12} to allow for long integration times. We have tested that starting with sonic white noise perturbation on $\bm{v}$ and without any inflow, the disc was hydrodynamically stable with $\alpha\sim 10^{-7}$ after a few local orbits, as expected with our density profile (\ref{eq:density_profile}). The resolution of each run was $(N_R,N_\theta)=(512,512)$ for $\varepsilon=0.1$ or $(N_R,N_\theta)=(1024,1024)$ for $\varepsilon=0.05,$ which corresponds to a locally isotropic resolution of eight points per $H$. We checked that doubling the resolution of run S7-1 did not quantitatively affect our results. We are therefore confident that our simulations are numerically converged. 

\subsection{Units, diagnostics, and notations}
Length units are given in $a.u.$, surface densities in $\mathrm{g.cm}^{-2}$, time units in orbital periods at $100\,a.u.=\td$ (or equivalently in units of $10^3\,\mathrm{yr}$), and accretion rates in $M_\odot.\mathrm{yr}^{-1}$.

In the following, several diagnostics are used to measure the behaviour of the disc coupled to the inflow. We first introduce two averages, an azimuthal average $\overline{\,\,\cdot\,\,}$ and a temporal average $\langle \cdot\rangle$. We start the temporal average from $t=10\,\td$ to allow the system to relax from the initial condition (this corresponds to $10^4$ orbits at the inner boundary).

These averaging procedures allow us to define fluctuating quantities $\delta\bm{v}=\bm{v}-\overline{\Sigma \bm{v}}/\overline{\Sigma}$. The first diagnostic is the \cite{SS73} $\alpha$ parameter 
$\alpha\equiv\overline{\Sigma \delta v_R\,\delta v_\theta}/\overline{P}$. We also use the disc radius $R_d$ defined as the location where the disc rotation velocity drops below $90\%$ of the Keplerian velocity  $v_\mathrm{K}$. This definition might look rather arbitrary, but it nevertheless leads to a well-defined disc radius since the rotation velocity drops very rapidly with radius at the transition region between the disc and the inflow. Each simulation is integrated for $3\times 10^4$ orbits at the inner boundary, which corresponds to $30\,\td$.

Our runs are labelled "XY-Z" where X=S or A denotes symmetric ($\tinj=2\pi$) or asymmetric ($\tinj=\pi/2$) inflows, Y is the accretion rate at $\Rout$ and Z is the specific angular momentum $\Lout$. Our simulations all have $\varepsilon=0.1$, except for runs labelled "thin", which assume $\varepsilon=0.05$. 
\section{Results}
We first concentrate on symmetric inflows. We then explore the impact of asymmetric inflows and of the disc thickness in a second part. 
\begin{figure*}
   \centering
   \hspace*{-1.9cm}\includegraphics[width=1.2\linewidth]{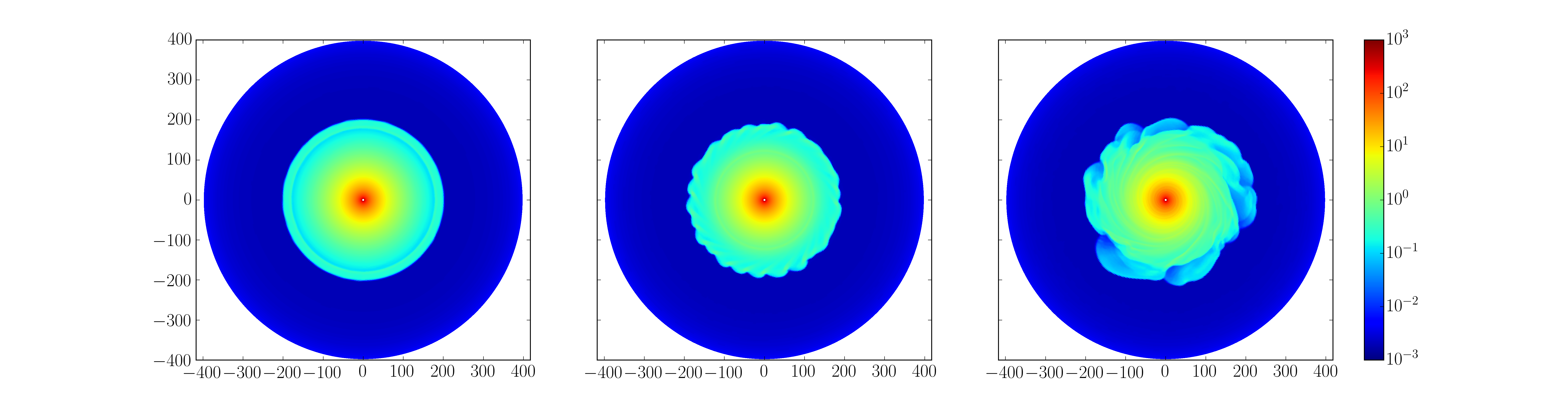}
   \vspace{-6mm}
   \caption{Density maps as a function of time for run S7-1. From left to right: $t=2\td$, $t=2.8\,\td$, $t=6\,\td$.}
              \label{fig:snap}%
\end{figure*}

\begin{table}
\centering
\setlength{\tabcolsep}{8pt}
\begin{tabular}{ l cccccc}
\hline
Run & $\Mout$ & $\Lout$ & $\alpha(\Rd/2)$ &$\alpha(\Rd/10)$ & $R_\mathrm{d}$\\
\hline
\hline
S8-1       &  $10^{-8}$ & $1$ & $1.6\times 10^{-3}$ & $5.3\times 10^{-5}$ & $255$        \\      
S7-1       &  $10^{-7}$ & $1$ & $4.5\times 10^{-3}$ &$1.4\times 10^{-4}$ & $199$        \\      
S6-1       &  $10^{-6}$ & $1$ & $1.0\times 10^{-2}$ &$3.5\times 10^{-4}$ & $167$        \\      
A7-1       &  $10^{-7}$ & $1$ & $3.5\times 10^{-3}$&$3.8\times 10^{-4}$ & $184$                        \\      
\hline
S8-1-thin       &  $10^{-8}$ & $1$ & $3.8\times 10^{-4}$ &$1.3\times 10^{-5}$ & $225$        \\      
S7-1-thin       &  $10^{-7}$ & $1$ &  $8.3\times 10^{-4}$ &$3.4\times 10^{-5}$ & $164$        \\      
S6-1-thin       &  $10^{-6}$ & $1$ &  $1.5\times 10^{-3}$ &$1.1\times 10^{-4}$ & $135$        \\      
\hline
S9-0            &  $10^{-9}$ & $0$ &  $8.1\times 10^{-4}$ &$2.8\times 10^{-5}$ & $264$                        \\      
S8-0            &  $10^{-8}$ & $0$ &  $2.3\times 10^{-3}$ &$1.0\times 10^{-4}$ & $177$                        \\      
S7-0            &  $10^{-7}$ & $0$ &  $4.2\times 10^{-3}$ &$3.0\times 10^{-4}$ & $91$                 \\      
A7-0       &  $10^{-7}$ & $0$ &  $4.5\times 10^{-3}$ &$5.0\times 10^{-4}$ & $90$                 \\      
\hline
\end{tabular}

\vspace{1mm}
\caption{\label{tab:runs}List of simulations. $\Mout$ is measured in $M_\odot.\mathrm{yr}^{-1}$, and $\Lout$ in units of specific angular momentum of a Keplerian disc at $R=100$: $\mathcal{L}_{100}$. } 
\end{table}

\subsection{Symmetric accretion}
\subsubsection{Accretion shock}
\begin{figure}
\vspace{-5mm}
   \centering
   \includegraphics[width=0.9\linewidth]{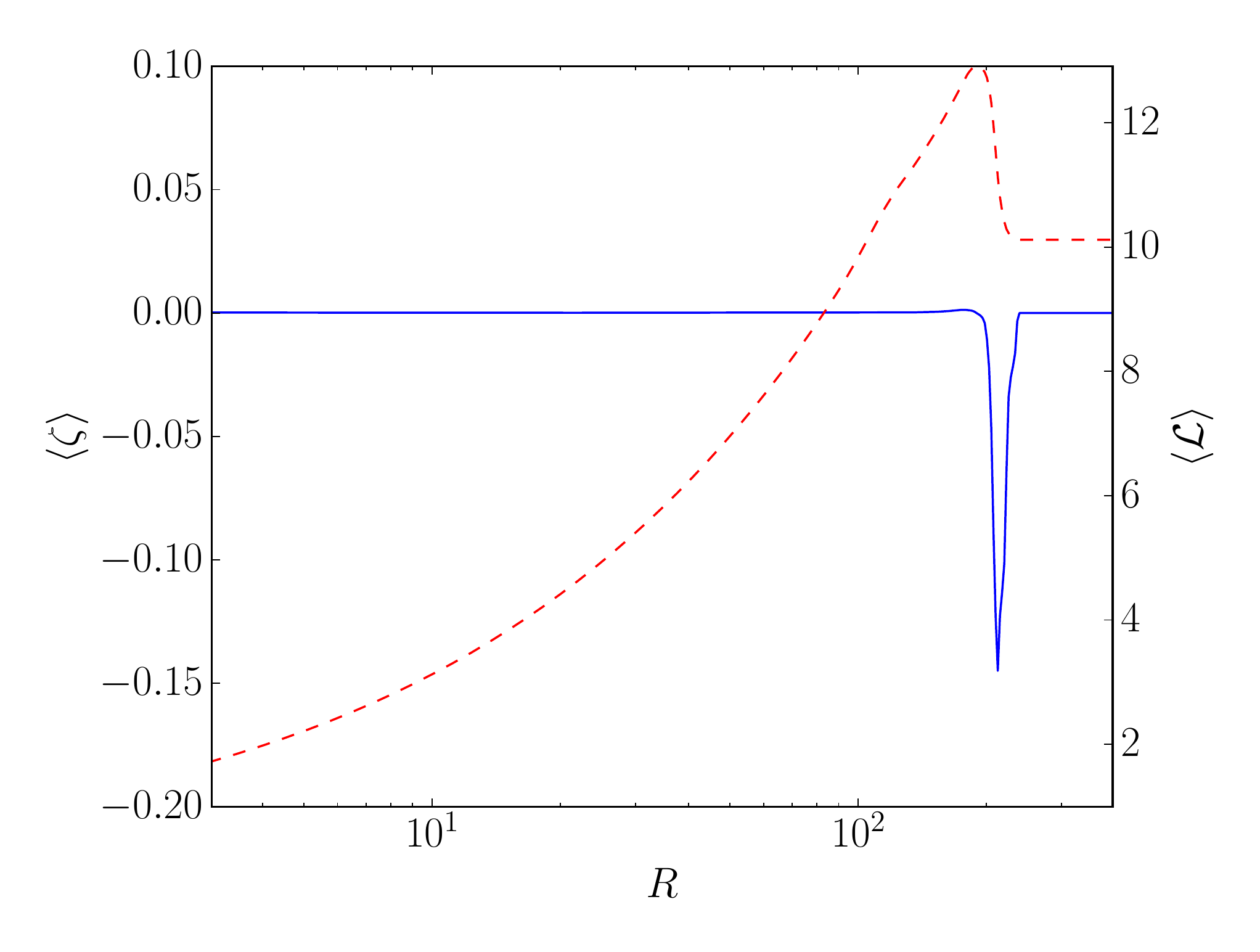}
   \vspace{-5mm}
   \caption{Averaged potential vorticity $\langle \zeta \rangle$ (plain blue) and specific angular momentum $\mathcal{L}$ (red dashed) of run S7-1. Note the sharp negative angular momentum gradient associated with a strong potential vorticity peak at $R\simeq 200$.}
   \vspace{-4mm}
              \label{fig:pv}%
\end{figure}

The initial evolution of the symmetric accretion run S7-1 is presented in Fig.~\ref{fig:snap}. The falling material initially forms an accretion shock that propagates inward. At $t\simeq2\,\td$, the shock stalls at $R=\Rd$. The stationary shock then develops an instability, which appears as a short-wavelength non-axisymmetric ondulation of the shock front ($t=2.8\,\td$). This instability then quickly saturates and generates strong spiral waves that propagate inward on the long term ($t>3.5\td$).

The origin of this instability is tightly linked to the structure of the shock. Spirals are continuously produced from $R\lesssim \Rd$, and the instability survives for the entire duration of the simulation. The instability mechanism becomes clear when the average specific angular momentum profile $\langle \mathcal{L}\rangle$ and the average potential vorticity $\langle \zeta \rangle\equiv \langle (2 \Omega+R\Omega')/\overline{\Sigma}\rangle$ are plotted as a function of radius (Fig.~\ref{fig:pv}). We find that at the location of the stationary shock a strong velocity gradient develops, which leads to an extremely peaked $\zeta$. This sharp structure appears because the specific angular momentum of the falling material is lower than that of the disc material at $R=200$. A potential vorticity layer is therefore unavoidable if matter keeps falling onto the disc.

As is well known, such a hydrodynamical structure naturally leads to a Kelvin-Helmholtz instability (KHI, also known as Rossby wave instability, or RWI, in the astrophysical context\footnote{In our case, the RWI is not due to a density bump but to a strong and localised shear. Nevertheless, the general RWI criterion, $\zeta^{-1}$ having a maximum \citep{L99}, is satisfied.}). We
note, however, that the shear rate is locally so strong that it also locally violates the Rayleigh criterion: in the shock region, the specific angular momentum is decreasing outward (Fig.~\ref{fig:pv}). As stated above, this configuration is unavoidable given that the inflow has a lower angular momentum than the disc. The instability driving the spiral is therefore a mixture of KHI (RWI) and Rayleigh centrifugal instability. It has a growth rate close to the orbital frequency at $R_\mathrm{d}$ and is sustained, for reasons stated above, during the entire duration of the simulation. 

\begin{figure}
   \centering
   \vspace{-5mm}
   \includegraphics[width=0.9\linewidth]{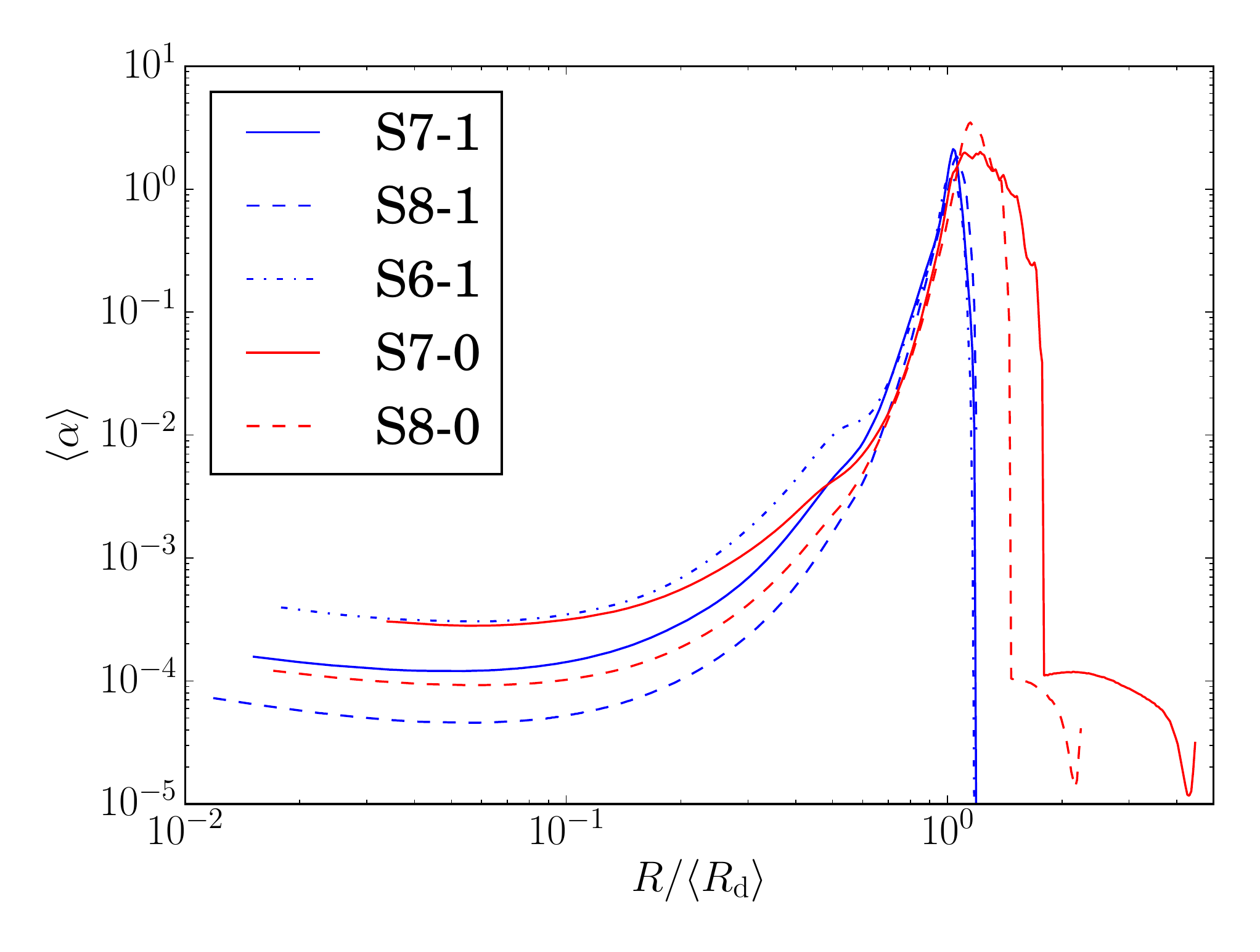}
   \vspace{-5mm}
   \caption{Profiles of $\alpha$ as a function of radius, measured from $\Rd$. The spiral stress increases with increasing $\Mout$ and with decreasing initial angular momentum $\Lout$. }
              \label{fig:alpha_sym}%
              \vspace{-1mm}
\end{figure}
\subsubsection{Angular momentum transport}
Spiral waves generate a positive Reynolds stress through the entire disc, which we quantify as an $\alpha$ parameter (Fig.~\ref{fig:alpha_sym}). We find that these spirals produce a very strong transport close to $R\simeq \Rd$ with $\alpha\simeq 1$. They propagate inward from $\Rd$ and dissipate through shocks, reaching $\alpha\gtrsim 10^{-3}$ at $\Rd/2$ and $\alpha\gtrsim10^{-4}$ at $\Rd/10$.  A generation region ($0.6\Rd<R<\Rd$) can be defined where the inflow mixes with the disc, spiral modes are excited, and $\alpha$ values are similar for all of our simulations. Likewise, there
is a propagation region ($R<0.6\Rd$) where the disc density structure is unaffected by the inflow except for propagating spiral waves.

\begin{figure}
   \centering
   \includegraphics[width=0.9\linewidth]{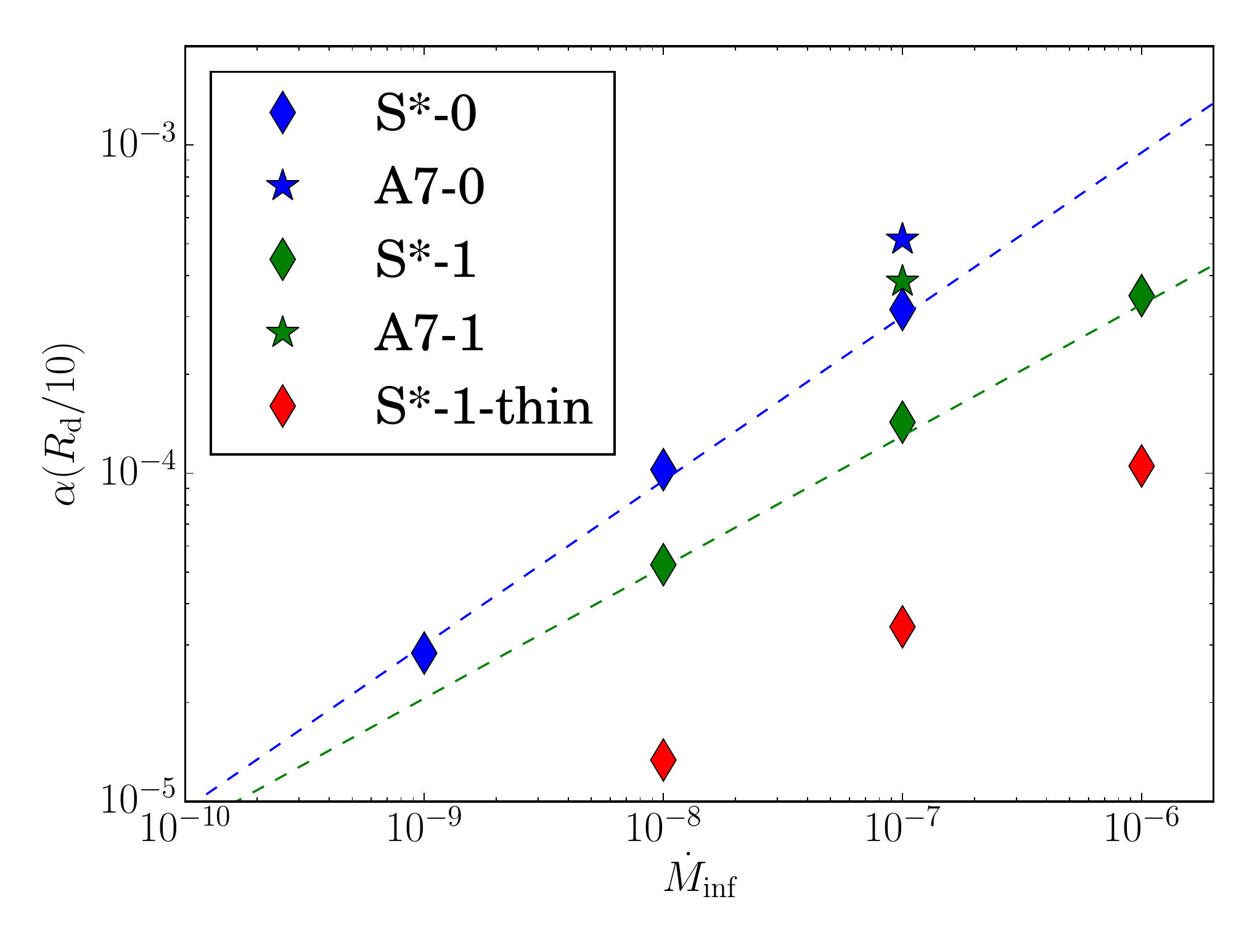}
   \vspace{-5mm}
   \caption{$\alpha$ measured at $R=\Rd/10$ as a function of $\Mout$. Colours represent different $\Lout$ or $H/R$, diamonds correspond to symmetric inflow, and stars correspond to asymmetric accretion. The dashed line are fits given by Eq. (\ref{eq:predictor}).}
              \label{fig:mdot_alpha}%
\end{figure}

We see that $\alpha$ in the generation region is not affected by $\Mout$, but $\alpha$ in the propagation region ($R<0.6\Rd$) clearly is. Moreover, for our strongest $\Mout$ (run S6-1 for instance), we observe a bump of transport for $R\simeq 0.6\Rd$, directly at the transition between these two regions. This is because the mass of the generation region has increased significantly due to the inflow while that of the propagation region has not, forming a jump in surface density at the interface. As a result, the generation region has a higher inertia and excites stronger waves at the interface with the propagation region.  We note that smaller angular momentum in the inflow also leads to somewhat stronger spirals. These findings can be summarised by representing $\alpha$ at $R=\Rd/10$ as a function of $\Mout$ (Fig.~\ref{fig:mdot_alpha}).

From these results, we can estimate that in the symmetric inflow case
\begin{align}
\label{eq:predictor}
\alpha(\Rd/10)&\simeq \alpha_0 \, \Bigg(\frac{\Mout}{10^{-7}\,M_\odot\cdot\mathrm{yr}^{-1}}\Bigg)^{\gamma}.
\end{align}
For inflows with significant rotation ($\Lout=1$) we have $\alpha_0=1.3\times 10^{-4}$ and $\gamma=0.4,$ whereas for inflows without initial rotation we obtain $\alpha_0=3.0\times 10^{-4}$ and $\gamma=0.5$. These values constitute a \emph{\textup{lower bound}} for the average alpha found in the disc. Typical values of $\alpha$ at $R= \Rd/2$ are 10 to 100 times higher than these estimates (Table~\ref{tab:runs}).

As is well known, standard accretion disc theory allows deriving a theoretical mass accretion rate from the profile of $\alpha$ computed above (\citet{BP99}, Eq. 28). 
We have checked that the measured accretion rates are consistent with this theoretical expectation, provided that $\dot{M}$ is computed on long enough averages (typically more than ten local orbital periods). This shows that standard accretion disc theory also applies to spiral-driven transport of angular momentum. This result does not imply, however, that ``turbulent'' heating follows \cite{SS73} $\alpha$-disc model \citep{BP99}. This point cannot be tested in our model since we used a locally isothermal equation of state.

\subsection{Parameter survey}

Since spiral waves are primarily sound waves, a dependence on the sound speed and therefore on $H/R=\varepsilon$ is expected. To test this effect, we reduce $H/R$ by a factor 2 with $\varepsilon=0.05$.  These simulations, labelled ``thin'', show that the amplitude of the spirals is significantly reduced when $H/R$ is reduced. We find a typical attenuation of a factor 3 between $\varepsilon=0.1$ and $\varepsilon=0.05$. This is similar to the case of tidally induced spiral waves \citep{SPL94}, despite an excitation mechanism different from ours. This similitude indicates that the wave propagation is most likely mainly responsible for this dependence.

All the results presented above were obtained in a situation where gas is falling onto the disc in an axisymmetric way. To test the effect of this assumed symmetry on the results, we consider the case of accretion coming from a narrower stream (runs labelled A*). We find that the disc shape is largely unaffected for the inflow rates we used and produces spiral waves similar to the one found in the symmetric case (Fig.~\ref{fig:snap}). In particular, we do not find any non-axisymmetric feature close to the location where the stream hits the disc. In addition, the azimuthally averaged profiles of $\langle \zeta\rangle$ and $\langle \mathcal{L}\rangle$ are very similar to the one found in the symmetric accretion case. The \emph{\textup{average}} flow is still unstable to the same instability as in the symmetric case. It is therefore expected to find that asymmetric accretion leads to $\alpha(R)$ profiles that are qualitatively comparable to those found in the symmetric case.

By comparing $\alpha(r)$ obtained in the asymmetric case to the symmetric case, we find that asymmetric accretion in general drives a slightly more efficient angular momentum transport, probably thanks to the excitation of low $m$ modes of the instability by the stream that propagates more easily. At $R=\Rd/10$, we obtain an $\alpha$ about two to four times larger than the one obtained from symmetric accretion simulations (Fig.~\ref{fig:mdot_alpha}). This suggests that the symmetric accretion scenario given by Eq.~\ref{eq:predictor} constitutes a lower bound to spiral-driven angular momentum transport.

\section{Discussion}
We explored the impact of an external inflow on a protoplanetary disc using 2D hydrodynamical simulations. We found that this interaction leads to the generation of strong spiral waves that propagate on long distances (typically down to radii smaller that $\Rd/10$). The resulting $\alpha$ at $\Rd/2$ are larger than $10^{-3}$ and reach a few $10^{-4}$ at $\Rd/10$ for inflow mass rates higher than $10^{-8}\,M_\odot.\mathrm{yr}^{-1}$. This stress varies strongly with $\Mout$ and $H/R,$ but is not significantly affected by the geometry of the inflow.

As shown in the introduction, this study is similar in nature to the work of \cite{BHZ15}. However, several important differences should be emphasised. First, our simulations only assumed a radial inflow and no vertical inflow. Second, the angular momentum of the falling material is always smaller than that of the outer disc so that a radial shear layer is always present, a possibility only explored in one simulation by \cite{BHZ15}. Because of these differences, we do not observe the formation of vortices, most probably because the instability at $\Rd$ involves a mixture of the RWI and of Rayleigh centrifugal instability thanks to the strong shear layer. We also obtained higher $\alpha$ values than did \cite{BHZ15} in the generation region, probably due to the same shear layer.

Finally, spiral waves such as we discussed here are now detectable with infrared continuum observations. Several authors have already reported direct observations of spiral patterns at the outer edge of evolved protostellar discs \citep[e.g.][]{MG12,BJ15}. Whether an external inflow might be at the origin of these spiral pattern is an open question that we defer to a future publication.

\begin{acknowledgements}
The computations presented here were performed using the Froggy platform of the CIMENT infrastructure (https://ciment.ujf-grenoble.fr). SF acknowledges funding from the
European Research Council under the European Union's Seventh Framework Programme (FP7/2007-2013) / ERC Grant agreement n 258729.
\end{acknowledgements}
\bibliographystyle{aa}
\bibliography{biblio}
\end{document}